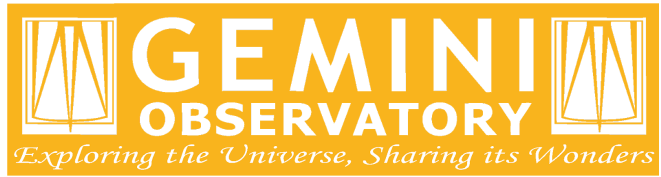

# Strategic Scientific Plan for Gemini Observatory

July 2, 2019

Prepared by: Gemini Observatory

Approved by the Gemini Board of Directors at their May 2019 meeting.

# Table of Contents

# Executive Summary

This document presents the Strategic Scientific Plan (SSP) for the direction and scientific activities of Gemini Observatory through the decade of the 2020s. The overarching goal is to ensure that Gemini best serves its international user community by remaining at the forefront of ground-based astronomical research throughout the coming decade. The development of this SSP was inspired by the document *Beyond 2021: A Strategic Vision for Gemini Observatory*, approved by the Gemini Board in May 2017. That document incorporated community input to identify a set of general principles for Gemini's future direction and made several broad recommendations based on those principles. The current document is aligned with those principles and recommendations, but proposes specific actionable items and associated strategies. The actionable items fall into three general categories: (1) preserving Gemini's current facilities and strengths; (2) developing instrumentation and software systems to enable new scientific capabilities that build on those strengths; (3) strategizing how visiting instruments can deliver additional valuable capabilities without overly taxing operational resources.

Gemini Observatory will continue to place a high priority on maintaining its current instruments and infrastructure while dealing proactively with the risk of obsolescence in both its hardware and software systems; several possible upgrades to the existing suite of instruments are proposed. In addition, Gemini should continue to offer its users a diverse set of observing modes and routes to obtaining telescope time, adjusting the balance according to the demand. Steps should be taken to guard against adverse impacts on the execution and completion rates of non-transient science programs in the wake of increasing numbers of targets of opportunity.

Gemini will implement efficiency improvements as part of the ongoing Observatory Control System upgrade and develop new systems to optimize its rapid response capabilities in order to be ready for the expected increase of transient follow-up proposals with the advent of the Large Synoptic Survey Telescope and other Time-Domain surveys. The eight-channel SCORPIO imaging spectrometer will be central to the time-domain follow-up capabilities at Gemini South. Development of robust, modular data reduction pipelines, both as an essential element for time-domain observations and to facilitate the attainment and publication of scientific results for all Gemini users, should continue with high priority.

In addition, the renewed emphasis on adaptive optics (AO), supported by the Gemini in the Era of Multi-Messenger Astronomy (GEMMA) award from the US National Science Foundation, will continue with the development of a new Multi-Conjugate AO (MCAO) system at Gemini North and a robust new real-time controller platform for all Gemini's AO systems. Planned improvements to make GeMS, the Gemini South MCAO system, more efficient and easier to schedule should also be implemented. Detailed planning for an Adaptive Secondary Mirror that will enable a Ground-Layer AO system at Gemini North will begin as soon as development resources become available. This SSP also proposes strategies to maximize the return and reduce the operational impact of Gemini's highly popular Visiting Instrument Program by focusing on innovative instruments that will provide new capabilities for broad segments of the user community, including instruments to make use of the new AO system.

The penultimate section of this SSP provides a high-level timeline (schematically illustrated in one figure) for the main developments discussed in the SSP. The landscape of ground-based astronomy is changing rapidly, and the schedule is consequently ambitious, but in light of the GEMMA funding, it is achievable with appropriate planning and prioritization of effort. Lists of milestones are given for gauging progress. As these milestones are reached and new instruments become available, some current instruments will need to be retired; the SSP makes recommendations in this regard. The final section concludes by reemphasizing the importance of a strong partnership committed to the needs of all members.

> *"What's past is prologue, what to come*
> *In yours and my discharge."*
>     — *W. Shakespeare, The Tempest*

## 1 Introduction: The Challenges on the Horizon

For nearly two decades, Gemini Observatory has sought to advance our knowledge and understanding of the Universe by enabling scientific study of the entire sky with its twin 8.1-m telescopes in Hawaii and Chile, outfitted with a suite of advanced astronomical instrumentation. Although the landscape of astronomy is ever changing and will see even more dramatic evolution in the near future, Gemini's core mission remains the same. To remain faithful to this mission, Gemini must continually strive to advance its instrumentation, operations, and user support to best serve the needs of its international user community. As defined by the Gemini Board, this Strategic Scientific Plan (SSP) is intended to be a "detailed roadmap for reaching a preferred future." Thus, it should serve as a guide for the Observatory along the path to fulfilling its mission throughout the course of the coming decade and advancing its position at the cutting edge of astronomical research.

The strategies described in this SSP are aligned with the recommendations put forward in the document [*Beyond 2021: A Strategic Vision for Gemini Observatory*](#) (hereafter *Beyond 2021*), approved by the Gemini Board at their May 2017 meeting. *Beyond 2021* was developed by a subcommittee of the Board by analyzing the results of an extensive 2015 survey of Gemini users that was completed by over 500 respondents from across the full partnership, as well as through consultation with the Gemini Science and Technology Advisory Committee (STAC) and the Users' Committee for Gemini (UCG). The respondents on the survey were asked to rate their level of agreement with wide-ranging "principles" and "scenarios" for Gemini's future evolution. The subcommittee then distilled those responses into several broad recommendations for Gemini's direction "beyond 2021." The recommendations include: exploiting Gemini's geographic locations and operational model to be the premiere follow-up facility for targets identified by the Large Synoptic Survey Telescope (LSST) while preserving a significant fraction of time for "Principal Investigator-driven science," being the premiere hosting facility for ambitious "facility-class" visiting instruments, and allowing the two Gemini telescopes to diverge in scientifically productive ways. By PI-driven science, the Board specifically meant programs targeting non-transient science targets, as opposed to time-domain follow-up observations.

*Beyond 2021* adds, "Synergies with other facilities (e.g., JWST) should also be exploited where they can be identified, and where the particular aspects of Gemini… can be exploited to good effect." This reflects the agreement expressed in the user survey with the principle that Gemini should build on its strengths in agile queue scheduling and its investment in facility adaptive optics (AO). In short, *Beyond 2021* calls for "exploring independent evolution" of the two telescopes, involving possible specialization of one or both telescopes in order to maximize synergies with other major facilities, while building on the existing strengths of Gemini and preserving diverse science. These considerations have guided the development of this Strategic Scientific Plan for Gemini Observatory.

While seeking to facilitate a broad range of science, this SSP also draws inspiration from the missions of the two groundbreaking new facilities for optical-infrared astronomy that will begin operations near the start of the coming decade: the aforementioned LSST and the James Webb Space Telescope (JWST). The former will revolutionize the young field of Time Domain Astronomy (TDA) through its wide-field, multi-band optical imaging survey, and the latter will provide stunningly high-resolution views of selected targets from the red end of the optical spectrum to the mid-infrared. To remain at the forefront of astronomical research, existing facilities must adapt their capabilities and operations to maximize synergies with these two transformative observatories on the horizon, as well as the pioneering facilities that have launched the new era of Multi-Messenger studies of our dynamic Universe.

Rising to the challenges ahead, Gemini is undertaking this multifaceted plan to upgrade its operations, instrumentation, and user support in order to provide the international Gemini community with essential scientific capabilities for the 2020s. These upgrades will benefit all users, regardless of their area of science or level of interest in other facilities, because the enhancements will make Gemini a more capable and efficient observatory.

The main developments discussed in this SSP include:

1. Preservation and enhancement of diverse science at Gemini, retaining multiple routes for obtaining observations and ensuring that new modes of operation do not adversely impact completion rates for any class of programs;
2. Operations upgrades to improve overall efficiency and to position Gemini as the premier 8m-class facility for discoveries in the Time Domain era;
3. A renaissance for facility AO at Gemini North;
4. Development of robust, automated data reduction pipelines for all Gemini facility instruments, with an immediate focus on robust automated reduction of long-slit spectroscopic data;
5. Strategizing the Visiting Instrument Program to focus on innovative instruments that will deliver new, high-demand capabilities to users.

In addition, there is the more basic necessity of maintaining and improving Gemini's suite of facility instruments and other essential infrastructure; this is addressed in the following section.

These are exciting times at Gemini. The international partnership has welcomed a new member with the formal signing by the Republic of Korea at the *Science and Evolution of Gemini* meeting in July 2018, and two institutions in Israel have signed limited-term partnership agreements, laying a foundation for more extensive future cooperation. Strategically important capabilities will be delivered in the coming years by facility instruments currently under development and by externally funded visitor-class instruments that are being built specifically for Gemini Observatory. In addition, the US National Science Foundation (NSF) has provided a major injection of supplemental funding in support of the Gemini in the Era of Multi-Messenger Astronomy ([GEMMA](#)) Program. The GEMMA award supports Gemini's efforts to achieve many of the core components of this SSP, to benefit the entire Gemini Community, as described in the following sections.

## 2 Maintaining Gemini's Instruments and Infrastructure

The preeminent priority for Gemini Observatory should be to ensure the most efficient and scientifically productive operations through regular maintenance, upgrades, and enhancements of its telescopes, infrastructure, and suite of instruments. Gemini is fortunate to have a dedicated and talented staff who daily maintain the Observatory's facilities and systems and nightly conduct its astronomical observations. The maintenance is done according to schedules determined by the needs of the various systems.

For example, the Acquisition and Guidance Unit (A&G) is serviced once annually, which requires downtime of the whole telescope. In the past this maintenance was carried out twice per year, but after implementation of several improvements, the system is now more reliable, making a yearly shutdown sufficient. In an operational observatory which runs a multi-instrument queue each night, opportunities for instrument maintenance can be difficult to accommodate. Previously, the A&G shutdowns were used to carry out instrument maintenance, but increasingly we are employing dedicated, pre-scheduled stand-downs per instrument to enable focused work to be done without the additional pressure of working on the A&G at the same time. On a longer timescale, mirror coatings are carried out approximately every 4-5 years; these major exercises typically involve staff from both Gemini sites and, increasingly in Chile, staff from the other local AURA facilities. Maintenance on shorter timescales is carried out via a work-order system, covering daily, weekly and monthly maintenance needs.

**The Longevity Program**

By the 2020s, both Gemini telescopes will be in their third decade of operations. We must therefore deal proactively to mitigate the threat of obsolescence in critical systems. As these systems continue to age throughout the coming decade, increased priority on this effort will be essential to maintain efficient operations and minimize downtime. This is especially important as more facility and visitor instruments are commissioned, because every instrument change entails some risk to the overall system.

Mitigation of these risks is the purpose of the Gemini Infrastructure Sustainability and Scientific Longevity Program. The Longevity Program, as it's known, seeks to identify, prioritize, and mitigate the most critical potential failures before they become a problem for nightly operations. It encompasses a broad range of projects where hardware, software, and processes may be updated proactively based on input from experts in all areas of the Observatory. The current mitigation plan is based on a proposal presented to NSF in 2017, which identified four major obsolescent systems, three IT-related required replacements, and a number of more minor issues identified by an internal audit that are being dealt with in a top-down order of priority. The Observatory Control System upgrade program, discussed in Section 4 below, is an example of critical higher-level software modernization.

**Telescope Infrastructure**

Three of the major issues identified in the Longevity Program relate to telescope infrastructure: (i) the obsolete real-time communications network by which the instruments and AO systems communicate with the secondary mirror, (ii) the secondary mirror control system itself, and (iii) motion control in the A&G unit. Technical solutions are being investigated for these issues.

Gemini is also exploring improvements in our mirror coating and primary mirror cooling procedures. Optical engineers from Gemini and GMT have jointly studied new multi-layer coating recipes that would greatly enhance the UV reflectivity, which currently drops steeply shortward of 400 nm. The study identified a promising, durable recipe that should yield reflectivity above 80% at 350 nm and above 90% longward of 400 nm (Schneider & Stupik 2018). It would require only minor modifications to the Gemini coating chamber. We will implement the physical changes in the GN coating plant in the near future. Test samples will be generated before moving on to actual telescope optics. Regarding the primary mirror cooling, active cooling was designed into the telescope but never commissioned; results from GPI (discussed below) indicate it is well worth implementing.

**Maintaining and Improving Gemini's Facility Instruments**

Many of Gemini's facility instruments are at least a decade old, and some have developed hardware issues that limit their observing modes. These issues are monitored by the Gemini staff, prioritized, and remedied when time and resources are available; current details can be found on the instrument status webpages. Here, we focus on significant upgrades that are needed to restore or enable key scientific capabilities to facility instruments. Some of the projects described in the paragraphs below are ongoing, while others have yet to be started; all should be given suitably high priority to move forward.

**GMOS-N** & **GMOS-S.** Both of these workhorse instruments have had their detectors replaced at least once, among other upgrades. To ensure their continued productivity into their third decades, additional upgrades will be required. For instance, the instruments could be commissioned for observations with GeMS at GMOS-S and the forthcoming GEMMA-funded MCAO system at GMOS-N. This would give factors of ~ 2 improvement in FWHM over natural seeing in red bandpasses (Hibon et al. 2016) with Strehl values reaching ~ 5% over a 2' field of view. Such a capability would create exciting possibilities for GMOS *r,i,z* imaging and spectroscopy at both sites.

Significant time and resources could be saved in the production and framing of GMOS masks with a more efficient mask-making system using state-of-the-art 3-D printers at both summits. This would enable pre-imaging and MOS observations within a single Fast Turnaround (see Section 3) cycle, as well as fast production of masks for spectroscopic characterization of sources in the vicinity of targets of opportunity identified in the LSST/multi-messenger era. Along the same lines of improving spectroscopic observing efficiency, a larger grating turret with positions for more gratings would simplify queue planning and improve completion rates in Bands 2 and 3. At the same time, the GMOS integral-field capabilities could be greatly enhanced by upgrading to larger IFUs covering wider areas. The IFUs are exchanged in the mask cassette mechanism; thus, adding a unit with coarser sampling over a wider area would enable a diverse array of new science without requiring modifications to the instrument.

With advances in CCD technology, Gemini should begin planning for the next upgrade of the GMOS detectors. For instance, e2v has developed new back-illuminated deep depletion 9K CCDs with 10 μm pixels and low read noise. A single such device would fill the GMOS focal plane, thereby obviating the longstanding need to "mind the gaps" when preparing GMOS observations and removing the associated complications in data reduction. The sampling of 0.05" $\text{pix}^{-1}$ would also be ideal for the AO-assisted imaging discussed above. Fully depleted devices of similarly large format may be on the horizon.

**FLAMINGOS-2.** The multi-object spectroscopy (MOS) mode of F-2 should be commissioned in the near future to provide Gemini with a powerful new near-IR MOS capability for diverse scientific applications. A longer term goal should be to commission F-2 long-slit and MOS modes with the wide-field AO capability of GeMS, or possibly to relocate F-2 to Gemini North for use with the new MCAO system there. In the longer term, a MOS-capable F-2 would be a very powerful instrument in combination with the ground-layer AO (GLAO) capability envisioned for Gemini North.

**GNIRS** & **NIFS.** As part of Gemini's Instrument Upgrade Program (IUP), a new IFU will be installed on GNIRS similar to the one that was destroyed in 2007, as well as a second high-resolution IFU that would be fed by the facility AO. This will require repair of the GNIRS short-red camera, which has a broken lens and has not been offered since 2014. The lens will be replaced in 2020, and the new IFUs should be commissioned in 2021. This upgrade will make GNIRS an alternative option to NIFS for AO-assisted integral field spectroscopy; NIFS may then be decommissioned, freeing up resources for other purposes.

**GeMS.** The GeMS natural guide star wavefront sensor (NGS2) upgrade will enable guiding on stars as faint as $r$ = 17 mag, improving sky coverage by a factor of three. Another important upgrade for GeMS is to restore the design multi-conjugate capability by replacing the third deformable mirror (DM) that was lost during commissioning. The optics and electronics of a replacement DM have passed acceptance testing; for more reliable GeMS performance, the new DM should be commissioned when resources allow. Gemini has also received funding to replace the GeMS real-time computer (RTC) as part of the GEMMA award from the NSF. The combination of the new, more stable Toptica laser, NGS2, the third DM, and a new RTC will make GeMS a much more robust and efficient instrument. Gemini is also working to make GeMS operable from the Base Facility and by a smaller team of observers; both of these changes will make it possible to schedule the runs more frequently. However, a major concern is the increasing number of planes flying over Cerro Pachón that disrupt laser operations repeatedly throughout the night; this situation requires external remediation.

**NIRI.** There is an ongoing IUP project to renovate one of the Gemini Polarization Modulator (GPOL) units and commission it as a visiting facility with NIRI. The feasibility report and the renovation plan are in progress and will need to be approved for the project to continue. If the upgrade proceeds as planned, GPOL will provide a powerful new capability for high-resolution polarimetric studies with NIRI of objects ranging from Solar System bodies to star formation regions to distant active galactic nuclei.

**GPI.** The Gemini Planet Imager is an extreme AO near-infrared coronographic imager and low-resolution spectrograph that can achieve very high contrast direct observations of exoplanets around bright stars. It has been in regular operation at Gemini South since 2014, and the Gemini Planet Imager Exoplanet Survey (GPIES; Macintosh et al. 2015) has observed over 530 stars in the southern sky. It is the newest of the current facility instruments, but with the completion of GPIES, the GPI team has been studying options for moving the instrument to Gemini North.

As noted by Macintosh et al. (2018), simply repeating the GPIES survey from Maunakea with the current instrument does not provide a very compelling science case. However, with upgrades, "GPI 2.0" could be made significantly more powerful for studies of the relative frequency of "cold-start" versus "hot-start" planets, properties of exoplanets around very young stars in the currently unreachable Taurus star-forming region, variability of exoplanets due to changing cloud patterns, and atmospheric properties via spectropolarimetry and high-resolution spectroscopy. An upgraded GPI 2.0 could also be useful for high-resolution imaging of asteroids and other solar system objects.

The science case is still being refined at this writing, and thus the precise requirements are not yet fully defined. However, the main upgrade possibilities include:

- Replacing GPI's Shack-Hartmann wavefront sensor (WFS) with a Pyramid WFS; tests indicate this would provide a gain of about 1.3 mag in sensitivity.

- Replacing GPI's current wave front sensor detector with an electron multiplying CCD (EMCCD) to provide another ~ 2 mag sensitivity gain and allow faster operations of the WFS for better AO correction on bright targets.

- Updating the real-time computer (RTC) to enable faster computation and AO correction speed, perhaps with control algorithms to predict atmospheric turbulence for even better correction.

- Improvements to the photometric calibration to achieve an accuracy of 1% for variability studies.
- Modifications to enable simultaneous spectroscopy and polarimetry.
- Installing a fiber feed to enable sending light to a separate high-resolution spectrograph.

Additional information on the GPI 2.0 plan is given by Chilcote et al. (2018). The upgrades would be done in a modular way to allow flexibility in case any descoping becomes necessary. Some risks have been identified by the GPI team for operations at GN, but these could be mitigated as part of the work related to the facility AO upgrade discussed in Section 5 below. In fact, the two efforts would be quite complementary, as the GPI team have developed expertise in characterizing sources of vibration and turbulence at GS and similar information in the North will be critical.

A somewhat surprising finding with regard to GPI's performance (in terms of contrast achieved) is that it is most strongly correlated with the temperature differential between the primary mirror and the outside air (Tallis et al. 2018). The performance is best when the temperature difference is small; the contrast worsens on average by a factor 2.5 for a temperature difference of 3° C. Thus, this finding adds urgency to the implementation of the active mirror cooling mentioned above.

**Facility Instruments in Development**

Two new and highly anticipated facility instruments, the Gemini High-Resolution Optical SpecTrograph (GHOST) and the Spectrograph and Camera for Observations of Rapid Phenomena in the Infrared and Optical (SCORPIO), are now in development for Gemini South. GHOST is a high-throughput, high-dispersion spectrograph with blue and red detectors that will provide simultaneous wavelength coverage from 363 to 950 nm. It has two selectable spectral resolution modes: standard-resolution with R ≈ 50,000 and high-resolution with R ≈ 75,000. SCORPIO is an 8-channel imager and spectrograph that will simultaneously cover the range from 385 nm to 2.35 μm. The eight independent arms in SCORPIO will allow the user to adjust exposure times in each bandpass independently. The negligible readout times of its detectors will enable observations at very high time resolution.

GHOST is scheduled for commissioning in 2020 and SCORPIO in 2022. Both of these instruments are anticipated to be workhorses at Gemini South throughout most of the 2020s. IGRINS-2 should be commissioned soon after, bringing a high-dispersion near-IR spectroscopic capability likely to Gemini North. The Observatory is committed to ensuring that these instruments perform to specifications and are commissioned with working data reduction pipelines. SCORPIO is specifically designed for rapid follow-up of transients identified by LSST. Additional plans for LSST follow-up are discussed in Section 4. Finally, as part of GEMMA, Gemini is also developing a major new facility AO system for Gemini North, generically known as GNAO; this is discussed in detail in Section 5.

Table 1 summarizes the plans and possibilities for the evolution of Gemini's instrument suite over the next five years; see Section 8 for a timeline and set of milestones for the new instruments. Given the abundance of instruments that will be available, some will necessarily be retired because of limited operational resources. This may entail difficult decisions. However, Gemini is committed to delivering high-demand capabilities to its user community, and this will guide the decisions to be made.

**Table 1. Gemini Facility/Resident Instrument Capability Outlook**

| Instrument | Avail | Vis | IR | Imager | Long Slit | IFU | MOS | hi-disp | AO now | AO 2025 | 2025 suite? |
|---|---|---|---|---|---|---|---|---|---|---|---|
| | | | | **Gemini North** | | | | | | | |
| ʻAlopeke | now | ✔ | | ✔ | | | | | spec | spec | Resident |
| GIRMOS | 2025 | | ✔ | ✔ | | ✔ | | | | ✔ | Resident |
| GMOS-N | now | ✔ | | ✔ | ✔ | ✔ | ✔ | | | | Facility |
| GNAOI[a] | 2024 | | ✔ | ✔ | | | | | | ✔ | Facility |
| GNIRS[b] | now | | ✔ | | ✔ | 2020 | | | ✔ | ? | Facility |
| GPI-2[c] | 2024[b] | | ✔ | ✔ | | ✔ | | | **XAO** | | Facility? |
| IGRINS-2 | 2023 | | ✔ | | ✔ | | | ✔ | | | Facility |
| MAROON-X | 2020 | ✔ | | | | | | ✔ | | | Resident |
| GRACES | now | ✔ | | | | | | ✔ | | | – |
| NIFS | now | | ✔ | | | ✔ | | | ✔ | | – |
| NIRI[d] | now | | ✔ | ✔ | | | | | ✔ | | (visitor?) |
| | | | | **Gemini South** | | | | | | | |
| FLAMINGOS-2 | now | | ✔ | ✔ | ✔ | | ✔ | | | ? | Facility[f] |
| GHOST | 2020 | ✔ | | | | ✔ | | ✔ | | | Facility |
| GMOS-S | now | ✔ | | ✔ | ✔ | ✔ | ✔ | | | | Facility |
| GSAOI[e] | now | | ✔ | ✔ | | | | | ✔ | ? | Facility?[e] |
| SCORPIO | 2023 | ✔ | ✔ | ✔ | ✔ | | | | | | Facility |
| Zorro | now | ✔ | | ✔ | | | | | spec | spec | Resident |
| GPI | now | | ✔ | ✔ | | ✔ | | | **XAO** | | – |

[a] GNAOI is the current generic name for an imager to be used with the new Gemini North AO system. Specifications are not yet fully defined; unlike GSAOI, it may include a natural seeing mode.

[b] GNIRS natural seeing and AO IFUs are to be commissioned in 2020; it is unclear if the AO mode will be adaptable to GNAO; if not, the AO mode may be lost, but GNIRS provides a unique spectral capability out to 5 μm that should be maintained.

[c] GPI-2, a significantly upgraded version of GPI, is not yet funded.

[d] NIRI will likely be decommissioned as a facility instrument but could find use as a visiting instrument if GNAOI does not include a wider field natural-seeing mode (to be determined).

[e] Plans remain to commission F-2 with GeMS, but the continuation of GeMS and GSAOI at Gemini South in the "time-domain era" will be contingent upon Observatory resources and Community demand.

[f] The continuation of F-2 as a facility instrument after the commissioning of SCORPIO will depend on the demand for MOS capabilities and the availability of Observatory resources; as discussed in the text, it could be relocated for use with GNAO and eventually a GLAO system.

## 3 Preservation of Diverse Science in an Evolving Observatory

The Gemini Board highlighted the continued importance of programs that study non-transient objects by making "Preservation of Principal Investigator science" one of the four Elements of the *Beyond 2021* Strategic Vision document, and recommended that "a significant fraction of the time on the telescopes should remain focused on Principal Investigator-driven science." As noted above, the intent of this statement was to support the continuation of diverse science programs targeting non-transient sources, rather than dedicating the Gemini telescopes to time-domain follow up. In light of this recommendation, broad instrument capabilities and a diversity of proposal options will remain essential for enabling the range of science pursued by Gemini's international user community.

Gemini currently provides multiple routes and observing modes for PIs in the participant communities to access time for their programs. The available options include:

- **Semesterly proposals**: Approximately two-thirds of the available science time, with regular deadlines every six months. The majority of these programs are carried out in Queue mode, with the exceptions being for block-scheduled visitor instruments and very rare Classical programs.

- **Fast Turnaround (FT)**: Up to 10% of the available science time; FT proposal deadlines occur at the end of each month, enabling expedited on-sky access and data delivery for small programs.

- **Large & Long Programs (LLPs)**: Up to 20% of the available time, with annual proposal deadlines during the *A* semester. The LLP option is intended for ambitious observing programs that require large time allocations and/or observations spanning multiple years. Beginning in 2018, Band 1 LLPs are guaranteed 80% completeness (barring major technical issues). The LLP time is "pooled" among the Gemini participants who choose to "opt in."

- **Director's Discretionary Time (DDT)**: up to 5% of the time, reserved for high-impact targets that require even more urgent observations than can be provided through the FT program.

- **Poor Weather Queue (PWQ)**: observing programs that can be carried out in very poor conditions (required image quality and cloud cover constraints are listed on the Gemini web pages) can be submitted at any time. If a PWQ proposal is technically feasible, it will be approved for execution when conditions are too poor for any program in the regular queue.

- **Subaru Exchange Time**: the Gemini and Subaru Observatories participate in a mutually beneficial time exchange agreement that makes the entire suite of instrumentation available at each observatory open to any astronomer within the other community. Proposals for exchange time are due at the usual semesterly deadlines. The typical time exchange in recent semesters has been five nights (50 hours).

For regular queue proposals and LLPs that have been awarded Band 1 status, PIs may request Priority Visitor (PV) mode for carrying out their observations. If approved for PV mode, the PI of a Band 1 queue program visits the Gemini North or South Base Facility and is trained to run the Gemini observing queue. The visiting PI may execute their program at any time that their targets are observable; for example, they may choose to observe their targets only during the best conditions, or they may relax their condition constraints in real time and go to their target if they determine that useful observations can be obtained. However, they cannot execute more time for their program than they have been awarded. In exchange for the priority that the visiting observers enjoy, they are required to execute the regular queue during periods when their targets are not observable, conditions are not adequate for their science, or when they have completed their own program allocation.

Target of Opportunity (ToO) observing mode is offered for queue observations of targets that cannot be specified in advance but which have well-defined external triggers. Examples include supernovae, gamma ray bursts, flare stars, tidal disruption events, or active bodies within the solar system. Proposals for the ToO mode may be made via the Semester, FT, or LLP proposal processes. These proposals may request either standard or rapid ToO (SToO or RToO, respectively) status, depending upon how quickly the observations must be carried out in order to be useful for the proposed science. Specifically, RToO observations must be executed within 24 hours of triggering and may interrupt the regular queue observing, while SToO targets simply become available for queue planning once they are triggered and can be scheduled any time after that.

**Challenge and Strategy**

The various proposal options and observing modes offer great flexibility, allowing PIs to tailor their proposals to the conditions, time, and cadence requirements of their science. The coming decade is expected to see a vast increase in the number of transient sources that can be followed up through ToO observations, and Gemini is preparing for this eventuality. However, the enthusiasm for rapid targeting of various classes of transient sources varies greatly across the Gemini partnership, as well as among different sub-communities within the individual partners. It is therefore essential that we preserve the quality and the diversity of the Gemini science programs that target non-transient sources, sometimes referred to as the "static sky."

Again, we note that *all Gemini follow-up of transient alerts will continue to be PI-led,* in the sense that they will be carried out by teams that have been awarded time through a competitive proposal-driven process. There is no current plan for a preordained amount of time to be reserved for a dedicated LSST transient follow-up campaign with the Gemini telescopes. However, the challenge is to preserve both the diversity and productivity of science programs that target non-transient phenomena, despite the expected increase in time-domain follow-up at Gemini.

To achieve this goal, we define the following requirements. First, PIs will continue to be able to propose for both "static sky" and ToO observations through the same variety of modes currently available. Second, observations carried out for the non-transient programs must not be negatively impacted by the expected increase in the number of transient follow-up observations. In particular, the non-transient program completion rates should not see any decline as a result of large amounts of LSST follow-up (at either Gemini site).

We will address these requirements in the following ways:

1. Gemini will continue to have semesterly calls for regular proposals, monthly FT calls, and annual LLP calls for PIs to propose programs targeting the static sky, standard ToO sources, and rapid ToO sources that need not make use of the network described below. The DDT and PWQ options for accessing Gemini time will also remain in place.

2. Gemini will participate with a community-driven, partner-dependent fraction of its time in a system known as the Astronomical Event Observatory Network (AEON), a network of telescopes that will be able to trigger automatically from a filtered stream of transient alerts, produced by a public "broker." (This system is described in more detail in the following section.)

3. Gemini operations software will be capable of "dynamic scheduling." This means that whenever a rapid ToO is triggered (either manually as done now or automatically by a program that gets its targets from an alert stream), the scheduler will automatically regenerate an optimized queue plan, including any interrupted target, if possible. If not possible (e.g., the target had set), the

       interrupted observation would be done with high priority on the subsequent night. Dynamic scheduling will also be used to regenerate the queue plan when conditions change, removing the need for numerous different queue plans each night.

4. Follow-up programs that require immediate interrupts (a small fraction even for RToO observations) will be required to include an "overhead" in their time request, which will be used to compensate fully for the time of any interrupted observations. This overhead will serve to deter PIs from insisting on more urgency than is actually warranted.

5. Triggers that do not require immediate observation would be folded into the existing queue by the scheduler, respecting the priority of the currently executing science program and other programs already in the nightly queue.

6. Simulations of the observing process will be used to anticipate the impact of varying amounts of ToO observations. Gemini will continue to monitor program completion trends to determine how non-transient observations are affected in practice and what limitations on the ToO observations are necessary to meet the requirement of minimal impact on completion rates.

In addition to dynamic scheduling, the ongoing Observatory Control System (OCS) software upgrade project will include other operational improvements to streamline the observation process. This means that a larger fraction of time each night will be spent collecting photons, improving overall program completeness. Following the start of the OCS upgrade project, Gemini was awarded additional funding as part of the GEMMA program for operations upgrades that will facilitate its participation in the AEON network, including automatic triggering of observations for programs that request it. Thus, the plan has the resources required to move forward.

# 4 Enabling Rapid Response and Dynamic Scheduling

The Gravitational Wave source GW170817, discovered by the Laser Interferometer Gravitational-Wave Observatory (LIGO), was a perfect example of the power of Multi-Messenger Astrophysics. The initial LIGO detection was followed 1.7 sec later by a short duration gamma ray burst detected by the Fermi and INTEGRAL satellites, then hours later by the discovery of an optical counterpart with ground-based medium-scale observatories and followed up spectroscopically on 8–meter class telescopes. X-ray and radio detections occurred in the days following. The comprehensive GW and electromagnetic (EM) data set confirmed that the source had all the signatures predicted for the cataclysmic merger of two neutron stars and demonstrated that such events likely produce the majority of heavy elements in the Universe. Moreover, the finding that ten Earth masses of the precious elements gold and platinum were created in this event captured the public imagination and created important outreach opportunities.

This GW170817 follow-up was conducted by multiple teams partially competing for the same resources with little overall coordination. Although the scientific return was substantial, with Gemini producing the most extensive near-IR data set available, the event demonstrated the need for better coordination in order to maximize future science. Until now, such highly compelling transient events have been rare and could be pursued by haphazard means. Starting in 2019, Advanced LIGO is expected to produce many more GW triggers, and by 2023, LSST will produce of order *ten million* transient alerts each night. This alert stream will include all objects that have appeared or brightened significantly since the previous observation of the same field. Other optical transient surveys such as Pan-STARRS, the Zwicky Transient Facility (ZTF), and SkyMapper will also generate alerts over the entire sky during this same period.

While only the small fraction of scientifically compelling LSST transient alerts will require following up, it is still a very large number in traditional terms. These alerts include rapidly moving sources that require orbital characterization in order to complete the census of potentially hazardous objects or to identify newly arrived interstellar voyagers such as 'Oumuamua; Type Ia supernovae for cosmological studies; core-collapse supernovae of all varieties for studying the diversity of stellar evolution; distant active galactic nuclei for understanding the process of supermassive black hole growth in galaxies; hypernovae candidates for constraining the rate of neutron star mergers; and even more exotic sources that defy attempts at classification and may represent new classes of phenomena. Thus, the science impacted by the coming generation of transient surveys involves everything from the Solar System to the epoch of reionization with enormous potential for probing the unknown.

**Challenge and Strategy**

The challenge is for Gemini to take advantage of its geographical co-location with LSST, its multi-site access to the sky from widely separated longitudes, its large aperture suitable for following up the faintest events, and its agile operational model, in order to maximize its discoveries in the vast new Time Domain opened by LSST and other surveys for the broad range of science areas listed above. An essential part of this challenge is knowing which targets to observe. More specifically, there must be a system in place to classify, select, prioritize, and follow up spectroscopically and/or photometrically the most scientifically compelling of the transient alerts using an optimized observational strategy.

Gemini has therefore joined a collaboration to create an efficient transient follow-up system known as the Astronomical Event Observatory Network ([AEON](AEON)). Besides Gemini, the AEON project currently involves the US National Optical Astronomy Observatory (NOAO), the Southern Astrophysical Research (SOAR) Telescope, and the Las Cumbres Global Telescope Network, a world-spanning network of small telescopes dedicated to observing bright transient phenomena. Other facilities may join in the future.

Observations on the AEON system will trigger from public alerts vetted by an "event broker," which monitors the public alert streams from various time-domain experiments and surveys. Science teams will create filters for the brokers that will select objects of interest which can then be managed and prioritized by target observation managers (TOMs). The TOMs are then used to send observation requests to the participating observatories on which the teams have been granted time and to monitor progress. Scheduling software will need to decide (based on conditions, priorities, and perhaps other considerations) if the requested target observation can be accommodated, and it will feed information back to the TOM. Figure 1 provides a schematic diagram of the AEON system modeled on the current Las Cumbres telescope network. More details are available on the AEON website.

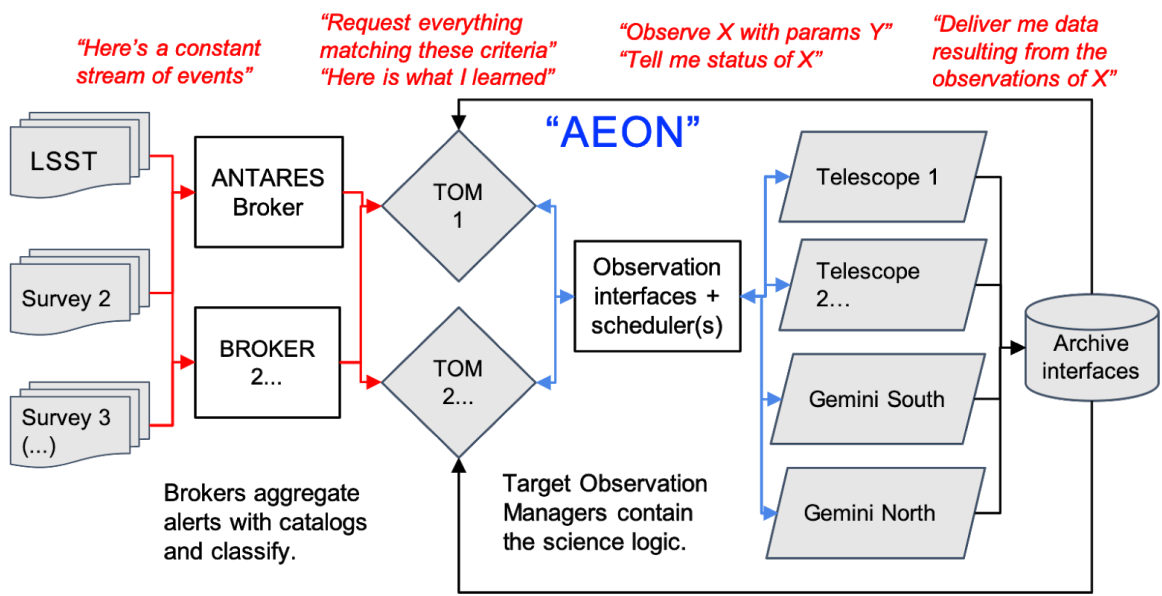

**Figure 1.** Schematic flow diagram for the AEON network designed for optimal follow-up of transient sources. The Gemini telescopes will participate in this network at levels to be determined by community demand. AEON encompasses the elements within the large box on the right, converting brokered transients into archived data. The possible "brokers" include the ANTARES system being developed by NOAO, other public brokers that can feed TOMs, and even individual PIs who manually create TOMs to trigger observations on targets of interest they have identified. Note that ToO programs on Gemini that do not flow through this network will remain an option.

Gemini's two telescopes will be the largest in the AEON system and thus responsible for characterizing the most challenging targets. Gemini's queue scheduling and agile operations model make it well suited for this effort; however, until now, ToO observations at Gemini have required manually triggering by PIs. In the era of LSST, follow-up will need to be conducted without the delays inherent in manual intervention, yet *without sacrificing other highly ranked science programs.* As discussed in the following section, robust data reduction pipelines are also needed for efficient automatic processing of the follow-up observations. Gemini is therefore undertaking major development work to ensure that its operations and data processing are optimally automated to schedule, execute, and process the observations of the most scientifically compelling transient targets uncovered by future sky surveys.

Because Gemini South shares the same sky and ambient conditions as LSST, these innovations are especially relevant for it. However, roughly 55% of the LSST transients will be reachable by Gemini North, and accounting for weather variations, 40% on average will be observable the same night. Of course, the longitude difference between the sites also enables longer continuous follow-up of sources visible from both hemispheres. Thus, Gemini North will also participate in AEON, though with proportionately less involvement in LSST follow-up and more in following up of northern surveys. Both telescopes, and the full user community, will benefit from the operations upgrades that will more efficiently schedule targets and protect the observations of non-transient sources as described in the previous section.

**Evolution of the Gemini Observatory Control System**

Gemini has begun a major software effort to modernize and streamline its Observatory Control System, the full suite of operations software used for observation preparation and execution. The OCS includes user applications such as the Gemini Phase 1 Tool (PIT) for proposal preparation and the Observing Tool (OT) for observation preparation. Despite many recent improvements to the OT (e.g., automatic guide star selection, automatic calibration settings, template observing sequences, incorporation of the integration time calculator, etc.), too many users find this software difficult to use. The design also makes it difficult to include the constraints needed for proper automated scheduling. At present, multiple static nightly observing plans are created manually for various sets of conditions, and it is cumbersome and inefficient for observers to update plans as new RToO observations arrive or to switch plans as conditions change. The expected increase in RToOs would lead to even more inefficiency.

With an eye on the requirements of the Time Domain era, Gemini is taking an opportunity to evaluate the operation model from first principles and determine which aspects will best serve the Observatory and its users in the coming decade. The OCS Upgrades Program is separate from the work on Gemini's participation in the AEON follow-up network, but the redesigned OCS is the platform by which Gemini will participate in transient follow up, and all other modes of observations. A requirement of the new OCS is that all the basic control services should be accessible through both manual user interfaces and APIs that define how the components operate within the system. These APIs will enable the TOMs used by follow-up science teams to query Gemini for telescope and instrument status, submit new observations, and determine the status of those observations. The new OCS will also facilitate the implementation of an automated scheduler and the interfaces between the observing database, the scheduler, and the sequencer that executes the observations.

Developing an automatic queue scheduler is a significant endeavor that will enable more efficient execution of all Gemini observations; it has implications for the OCS Upgrade Program, but the development is being carried out as part of the GEMMA Time Domain Astronomy project. While the Las Cumbres scheduler provides a useful guide, significant changes are required to use it for Gemini. A key difference is that observations on the Las Cumbres network cannot specify constraints on the conditions such as image quality and cloud cover that Gemini operations require. It is essential to maintain this capability in order to make optimal use of the best observing conditions and take full advantage of the Gemini telescopes' exquisite design.

Both the OCS Upgrade and Time Domain follow-up projects are in the advanced design stages and both are included in the budget (following the recent injection of funding by NSF for Time Domain follow up and AO development as part of the GEMMA program). The upgrades are intended to ensure that Gemini observations are optimally scheduled and efficiently executed, which also should benefit users studying the static sky. Another aspect of this, as discussed in the preceding section, is finding ways to balance the amount of time spent on transient versus non-transient sources; this should be driven primarily by demand from the user community.

# 5 Northern Focus on Adaptive Optics Capabilities

Gemini North's ALTAIR system was among the first facility AO laser guide star (LGS) systems in routine operation. In this context, "facility" means that the AO system feeds other instruments on the telescope. Although the use of an LGS greatly increases the sky coverage for an AO system with respect to what can be achieved with natural guide stars (NGSs), the well-corrected field of view (FoV) for a single conjugate AO (SCAO) system such as Altair is limited to around 10".

A number of techniques using multiple laser and/or natural guide stars have been proposed over the past decades to obtain correction over a larger FoV. A multi-conjugate AO system (MCAO), using multiple deformable mirrors to correct for atmospheric distortions occurring at different atmospheric levels, was first demonstrated on-sky at the VLT with the Multi-Conjugate Adaptive Optics Demonstrator in 2007 using multiple NGSs. The first facility MCAO system came online in 2011 with the Gemini Multi-conjugate System (GeMS) at Gemini South (Rigaut et al. 2014). GeMS uses a constellation of five laser guide stars, plus one to three natural guide stars, to achieve nearly diffraction-limited correction over a 1.4' field (the fully corrected field could be made as large as about 2' in diameter), much larger than is possible with an SCAO system.

**Experience from GeMS**

In addition to its agile operations model based on flexible queue scheduling, another of Gemini's key strengths is an opto-mechanical design optimized for superb image quality. The GeMS system at Gemini South, coupled with the Gemini South AO Imager (GSAOI), takes advantage of this strength to deliver nearly diffraction-limited K-band images with typically 0.085" resolution over a 1.4' field of view on an 8-m telescope. In late 2017, GeMS received a new, more stable Toptica laser that has greatly improved the reliability and efficiency of its performance.

Science results from GeMS/GSAOI include the deepest color-magnitude diagrams ever produced for obscured globular clusters within the bulge of the Milky Way (Saracino et al. 2016), providing new age and mass estimates for these clusters. In combination with earlier *Hubble* imaging, GeMS has determined the proper motion of the distant halo globular cluster Pyxis, finding that it is likely of extragalactic origin and yielding a lower limit on the total mass of the Milky Way (Fritz et al. 2017). Proper motion measurements by GeMS for high-velocity, subarcsecond-sized knots within the "molecular fingers" of the Orion OMC1 cloud core imply an explosive origin for the protostellar outflow (Bally et al. 2015). These same GeMS images were used to identify compact IR sources to search with ALMA at 1.3 mm for emission from protoplanetary disks (Eisner et al. 2016). A pilot GSAOI imaging survey of luminous infrared galaxies (LIRGs) discovered three core-collapse supernovae within dusty, crowded regions of intense star formation, indicating that the supernova rates within LIRGs are much higher than previously estimated, but most are missed as a result of dust obscuration and inadequate resolution (Kool et al. 2018). GeMS has also been used to study the near-IR morphologies of distant galaxies at spatial resolutions surpassing what *Hubble* can achieve at such wavelengths (Lacy et al. 2018).

In contrast, while Maunakea is the best existing astronomical site for AO performance, Gemini North lacks any wide-field AO capability and its aging single-conjugate ALTAIR AO system falls well short of fully exploiting the site's outstanding characteristics. In fact, no comparable MCAO system similar to GeMS exists in the northern hemisphere and there are no plans for one for another decade. The Thirty-Meter Telescope (TMT) will be equipped with the revolutionary NFIRAOS MCAO system, which will provide diffraction-limited AO imaging in the IR over a ~ 30" FoV on a 30m telescope, but it is unlikely to be available before the end of the 2020s.

**Challenge and Strategy**

Gemini's Science and Technology Advisory Committee has advocated for a world-class wide-field AO system for Gemini North, building on previous experience with GeMS. An advanced MCAO system at Gemini North would enable a broad range of studies focusing on stellar populations, supernova physics, outflows, proper motions, and galactic archaeology, as well as synergies with other facilities capable of high-resolution studies at longer and shorter wavelengths, such as ALMA and *Hubble*. These subjects have been explored to some degree by the GeMS studies referenced above. However, with the benefits of greater sky coverage from improved guide star sensors, more efficient queue-based operations, and the superior atmospheric conditions at Maunakea, such a system would be capable of a great variety of other science. Examples include monitoring long-period variables in galaxies as far as the Virgo cluster to constrain lifetimes on the thermally pulsating asymptotic giant branch in diverse populations, searches for intermediate-mass black holes from the internal motions of dense stellar systems, rest-frame optical studies of galaxy interactions at the peak of cosmic star formation activity at $z \sim 2$, and exploration of the earliest stages of galaxy formation via narrow-band imaging of Lyman-alpha emitters at redshifts where the emission falls between the atmospheric OH lines.

These topics are also key science drivers of JWST, but because of JWST's L2 orbit and limited range of pointing angles, it can only observe most targets for limited periods each year. As discussed in more detail in a related *Astro2020* science white paper (Blakeslee et al. 2019), this provides an excellent opportunity for synergy: an MCAO system on an 8m-class telescope on Maunakea would be the only facility able to study and monitor high-priority northern targets with a similar spatial resolution and FoV as NIRCAM on JWST. This ability is especially important for time-domain science involving candidate lensed supernovae, such as SN Refsdal (Kelly et al. 2015), for which additional lensed images may appear, depending on the lens model, when the target would be unobservable by JWST and when *Hubble* itself may no longer be operational. The challenge is to build such a system.

To enable Gemini to rise to this challenge, the NSF has awarded major new funding as part of the GEMMA Program to construct a cutting-edge MCAO system at Gemini North, referred to as GNAO. Operating on an exceptional site for AO performance, GNAO will establish a clear leadership position for Gemini North by enabling unique synergies with JWST for northern time-domain targets during periods when they are unobservable by JWST, as well as for simultaneous AO and spectroscopic follow-up of LSST targets visible from both Gemini sites. GNAO will be the cornerstone of a comprehensive Gemini North AO upgrade. It will build on the previous investment in GeMS but employ the latest technology for better performance in support of the next generation of AO-assisted instruments, such as the Gemini Infra-Red Multi-Object Spectrograph (GIRMOS; Sivanandam et al. 2018), an externally funded multiple integral-field unit spectrograph being built by a consortium of Canadian universities.

**Baseline Requirements for the Gemini North MCAO System**

GNAO will not be a simple copy of GeMS. Based on the scientific drivers above, and the Observatory's experience with GeMS, there are four top-level requirements:

- A two-arcminute field of view: this is the FoV of JWST/NIRCAM's two camera modules and is a physical limitation set by the size of the AO Fold mirror.

- With the advantage of Maunakea's atmospheric conditions, the system should deliver at least 30% Strehl over the full corrected field in *K* band for median seeing, with a goal of 50% Strehl.

- Point spread function (PSF) astrometry better than 1 mas, based on the community demand for

> astrometric programs with GeMS.

- The system should be queue operable by the standard Gemini night crew (one trained telescope operator and one trained queue observer), without any additional support, and thus available for observations on any night.

The technical development work to achieve these requirements is ongoing at the time of this writing, and the schedule is aggressive, with deployment planned for 2024.

To take full advantage of the wide-field high-resolution performance of the new MCAO system, Gemini North will require a modern, large-format IR imager with a pixel scale designed for optimal PSF sampling. The existing NIRI instrument will not be compatible with the f/32 design of GNAO. However, using existing development funds, the Observatory is planning to build a new Geminii North AO Imager (GNAOI) with capabilities similar to GSAOI, but with a single Hawaii 4RG detector filling the focal plane. Unlike GSAOI, it may also have a natural seeing mode with twice the field of view and a correspondingly coarser pixel scale. In addition, the previously mentioned GIRMOS multi-IFU instrument will include imaging capability, and could become a workhorse for both imaging and spectroscopy.

**Ground Layer AO**

Correction over an even wider field is possible with ground layer adaptive optics (GLAO) systems, which can provide a factor of about 2 to 3 improvement over seeing-limited observations over scales of 10' or more, but do not deliver diffraction-limited images in the near-infrared. GLAO systems generally involve a deformable, or adaptive, secondary mirror (ASM). ASMs are now coming online or being planned at several other 8m-class facilities (LBT, Magellan, MMT, Subaru, VLT), and new techniques are being developed to reduce their cost and complexity (e.g., Hinz et al. 2018).

The replacement of the current secondary mirror with an ASM at Gemini North would enable wide-field GLAO correction for improved imaging and spectroscopic performance for all non-AO instruments over the full wavelength range of the telescope, as no dichroics are involved. It would also benefit the performance of the AO instruments, including extreme AO systems such as GPI-2, as well as GNAO itself. Moreover, a GLAO system employing an ASM does not entail additional optical elements that would increase the thermal background. Therefore, for visiting instruments that seek to take advantage of conditions on Maunakea to maximize synergies with JWST in the mid-infrared, a GLAO system would efficiently enable diffraction-limited performance at wavelengths beyond 5 μm. Gemini will commission a full feasibility study for an ASM at Gemini North, as the potential scientific benefits are enormous.

# 6 Supporting Gemini Users with Modern Data Reduction Pipelines

Gemini is a queue-based observatory with a distinct partnership structure; as a result, it has developed a unique user support model. Support for Gemini users from proposal writing to data reduction and scientific analysis is provided through a combination of the National Gemini Offices (NGOs) and the Gemini staff. The NGOs provide initial support with program development and answer basic Helpdesk queries, and they provide essential feedback and interaction with their respective user communities. User interaction by the Gemini staff includes program development assistance from support scientists within the Gemini North and South Science Operations Departments and help with data reductions and related issues from members of the Science Users Support Department (SUSD). This distributed support model allows Gemini users to access a wide variety of experts both within their home communities and on the Observatory staff in order to ensure they have the tools, documentation, and assistance required to maximize the scientific return on their observations.

Another important function of the Gemini SUSD is the development and maintenance of data reduction software for Gemini facility instruments. Gemini currently provides data reduction software for all of its facility instruments. The purpose is to make the data amenable to scientific analysis by removing instrument, telescope, and atmospheric signatures (including sky/telluric features). For most instruments, this has been in the form of data reduction packages consisting of scripts that make use of the tasks within the Image Reduction and Analysis Facility (IRAF). However, there is no longer any development effort by NOAO or others on IRAF, and the package is rapidly growing obsolete. To deal with these issues, most major observatories are moving towards using more modern data reduction software. Gemini itself has begun a project to replace its IRAF data reduction packages with a Python-based platform that uses the [Astropy](Astropy) library and is aligned with its development.

**Challenge and Strategy**

In light of the efforts already discussed in this SSP to make Gemini the premier 8-m class optical facility for rapid characterization of transient events in the dawning Time Domain era, robust automated imaging and spectroscopic data reduction pipelines have become essential. The challenge for Gemini is to develop robust, modern data reduction tools that will allow easy incorporation into automated pipelines that will deliver science-quality data products to accelerate the process of garnering the critical information from one's observations and thus benefit the entire Gemini user community.

The SUSD is rising to this challenge by focusing its software development efforts on DRAGONS (Data Reduction for Astronomy at Gemini Observatory North and South), a modular, extensible data reduction platform being developed in Python (with no dependencies on IRAF). DRAGONS offers an automation system that simplifies pipelining the reduction of Gemini data, or of any other astronomical data that includes standard configuration information. DRAGONS will entirely replace the Gemini IRAF package for most facility instruments; data reduction software packages for all new facility instruments are required to use DRAGONS.

For Gemini instruments, the platform provides default parameters that work well for the majority of data sets, allows users to modify these defaults, and for some applications provides an interactive mode for optimization of the data reduction steps. DRAGONS software currently operates in Quality Assessment mode (QA) at the telescopes every night, returning sky condition metrics such as image quality (IQ), cloud cover (CC), and sky background (BG) as measured on any image, including acquisition images. This helps the observers to assess quickly the conditions in selecting among the various queue plans that are prepared for each night's observing.

Development efforts are on track to release a science-quality data reduction version of DRAGONS by 2020 that supports pipeline reductions for all active facility imagers. The next step will be GMOS long-slit spectroscopy (our most commonly used existing instrument mode), establishing the foundation of our Python spectroscopy software suite, the bulk of which will be reusable for Gemini's other spectroscopic instruments. This effort is complementary to the ongoing development of the DRAGONS-based SCORPIO pipeline, and the schedule will enable extensive testing of DRAGONS-based pipeline data products before the commencement of LSST operations. Long-term efforts include extending the interactive mode to a larger array of tasks. More details are included in the Milestones section at the end of this document.

DRAGONS will be central to maximizing the efficiency of Gemini's participation in the planned AEON follow-up system. Pipelines based on DRAGONS will automatically reduce all imaging and long-slit data and deliver the products to the Gemini Observatory Archive (GOA) for the PI to evaluate in real time so that decisions can be made as to whether or not to trigger additional observations. However, although DRAGONS development is essential to Gemini's role in AEON, and has received supplemental funding as a result, the development of robust, automated data reduction pipelines, based on modern software toolkits, has been advised with ever increasing urgency by the Gemini STAC, Users' Committee, and Operations Working Group, independent of any consideration of transient follow-up. This major initiative to revamp, modernize, and streamline Gemini data reduction will provide enormous benefits to the entire Gemini community of observers and anyone who uses GOA data products.

## 7 Strategizing the Visiting Instrument Program

Gemini Observatory has a current operations plan that includes four facility instruments plus a facility AO system (i.e., one that feeds other instruments) at each site, but the acquisition of new instruments is slow and requires large investments of development funds and other resources to oversee the instrument design and construction. The goal of Gemini's Visiting Instrument Program (VIP) is to provide users with a wider range of instrument capabilities for diverse science applications, while providing the instrument teams with opportunities to pursue their own science with instruments of their own design installed on a world-class 8-m telescope. In this way, Gemini can respond more quickly to new trends in research without the large overhead and cost involved in selecting a new facility instrument.

A visiting instrument is one built or procured by another organization and made available temporarily for use at Gemini via an agreement between the instrument's owner and Gemini Observatory through the VIP. These instruments are often originally built for other facilities, but we are also now receiving proposals for visiting instruments designed and built for Gemini from the start. To ensure the scientific viability of these visiting capabilities, our policy requires teams to go through the Gemini time allocation process to get time before using their instrument on Gemini. Instrument teams wishing to return for multiple semesters are additionally asked to offer the capability to the community of Gemini users as part of the regular call for proposals, with the instrument team committing to carrying out the observations in block scheduled observing periods.

In the *Beyond 2021* Strategic Vision, the Gemini Board advised the Observatory to prioritize ambitious visiting instruments that will deliver new high-demand capabilities to the users. The VIP strives to provide a balanced suite of cutting-edge capabilities, some with broad appeal and others that appeal to more specific science cases that may yield high profile results. Through regular user surveys, as well as interaction at AAS meetings and elsewhere, Gemini makes a concerted effort to gather feedback from the community on what visiting capabilities would be most desirable.

In order to attract a broad range of visiting capabilities, Gemini maintains an open Call for Visiting Instrument Proposals and publicizes the program through electronic and print media, as well as through our partner NGOs and visits to institutes with skilled instrumentation teams. The VIP then provides the support that the instrument teams need to bring their proposals to fruition. This includes assistance to groups who may have plans for a new instrument, but little knowledge of Gemini or experience writing proposals. The support can take the form of collaborative efforts to develop science cases, determine the level of community interest, and iterate on the details of the proposal itself. If a proposed visiting capability has the potential for high value as either a visiting or future facility instrument, some funding or development support may also be made available through the VIP. In this way, Gemini strives to provide access to a diverse and innovative suite of instrument capabilities for the user community.

Additionally, the VIP has begun to serve as an entry point for teams that ultimately wish to deliver a new facility instrument to Gemini, whether for their own science or to gain experience for future instrument projects on 30m-class telescopes. In such cases, the team can start by demonstrating the capabilities of the instrument and the demand on the part of Gemini's users. Once proven, Gemini and the team can mutually decide whether the instrument should continue on a visiting basis (with the team providing significant user support) or if it should be adopted as a facility instrument with the Observatory taking over all support and maintenance. IGRINS is a variation of this model: its high demand as a visiting instrument led to the plan to develop IGRINS-2 as a facility instrument.

**Challenge and Strategy**

The VIP has proven very popular with both instrument teams and the user community since it was established several years ago. Gemini typically receives two to three proposals per year from instrument teams, not all of which are well aligned with Gemini's strengths and mission, but some clearly are. The most extreme example of this is IGRINS, which became the first instrument to have a demand exceeding that of GMOS when it was offered as a visiting instrument at Gemini South for 2018A. Inspired by the successes of the VIP, the Gemini Board issued the following challenge in *Beyond 2021*: "Gemini should be viewed as the premiere hosting facility for visitor instruments whose scope and ambition may be comparable to that of the 'facility-class' instruments."

Thus, the challenge is to attract ambitious instruments that will bring new capabilities to Gemini and enjoy high demand. The basic strategy for meeting this challenge is straightforward:

- Continue to promote the VIP to instrument teams to attract the largest pool of proposals;
- Ensure allocation of the necessary resources to support the activities of the VIP at Gemini;
- Engage in detailed cost-benefit analyses in deciding which visiting instruments to accept and to continue to host.

The cost-benefit analysis must include considerations such as: (1) user demand, i.e., proposal pressure, (2) feedback on user surveys, (3) scientific impact of the publications based on the instrument, and (4) the amount of support required from the Observatory for the instrument. In consultation with the STAC, Gemini now considers all these factors in assessing proposals for visiting instruments and their ongoing performance.

Of course, the most straightforward way to determine community interest in an instrument is to track the proposal pressure. This is assessed each semester based on information received from the individual national TACs and during the Gemini ITAC process. The exceptionally high demand for IGRINS (35% of the available time in the one semester that it was offered) has already been noted, and it is not uncommon for other visiting instruments to have demands near 10% of the available time. If a visiting instrument does not receive sufficient demand, it will not be put on the telescope that semester. For the four semesters in 2017 and 2018, visiting instruments accounted for between 12% and 38% of the total demand at each telescope, which is higher than some facility instruments.

An example of a unique visiting capability at each site is provided by 'Alopeke and Zorro, which are twin low-noise, two-channel speckle imagers permanently mounted within the calibration ports at Gemini North and South, respectively. They are compact enough to fit alongside the calibration units without causing obstruction. 'Alopeke accounted for a respectable 7% of the demand in the first semester it was offered, and Zorro accounted for 12% in its first semester. The science enabled by the diffraction-limited speckle imaging capability, including studies of the binarity of exoplanet hosts, has resulted in several high-impact publications for Gemini (partly produced by the preceding speckle imager DSSI). Both 'Alopeke and Zorro have been used in multiple time-critical DD programs, which is unusual for a visiting instrument, but made possible by the permanent mounting and the instrument team's dedication to providing observing support and reduced data products.

One measure of the impact of the Visiting Instrument Program is the number of publications in refereed journals that result from these instruments. Gemini now keeps a record of publications using visiting instruments specifically, and this information will be used in the cost-benefit analyses mentioned above. So far the overall publication numbers are consistent with expectations based on the fraction of time allocated and the time elapsed since the data were taken. As of this writing, most of the data obtained from the large time allocation for IGRINS in 2018A are still in the analysis phase, but early publications have appeared (e.g., Tegler et al. 2019) and many in-progress results were highlighted at the "Science and Evolution of Gemini Observatory" meeting in July 2018.

**Outlook for Visiting Instruments**

The next two facility instruments to be delivered, GHOST and SCORPIO, will each bring powerful new capabilities to Gemini South in the near future. In contrast, Gemini North has not received a new facility instrument for more than a decade. However, the addition of the new state-of-the-art facility MCAO system at Gemini North, scheduled for commissioning in 2024, is attracting a broad range of visiting instrument ideas. In particular, the ambitious GIRMOS facility-class visiting instrument, originally slated for Gemini South to interface with the GeMS system, will likely go to Gemini North instead. The availability of MCAO at both Gemini North and South will provide more flexibility in choosing and scheduling visiting instruments, while reducing the pressure for community demand at Gemini South following the commissioning of GHOST and SCORPIO there. As the premier dry site in the northern hemisphere, Maunakea also provides opportunities for innovative teams to develop high-resolution synergies with JWST at wavelengths beyond the 2.5 μm cutoff of GIRMOS.

In planning for "facility-class" visiting instruments, it should be recognized that the VIP has been sustainable because of the requirement that the teams carry out the observations, both for their own science and for successful proposals from the community. Gemini staff are not trained on the operation of each visiting instrument; this ensures that the staff are not overburdened, as most visiting instruments are only mounted on the telescope for periods of days or weeks.

For the case of a long-term visiting instrument that brings new capabilities in high demand by the Gemini users, it may be in the Observatory's interest to waive this requirement. This could apply to instruments that are at Gemini for a continuous period exceeding one semester and provide capabilities that are in high demand by the community. A potential example of this is IGRINS, the high-resolution infrared spectrograph that enjoyed high demand during the semester it was at Gemini South. If the community remains highly receptive, and the instrument returns for a year or more, it may be advantageous to relax the guidelines and treat it as a facility instrument during the remainder of its stay, since it would be unlikely that the instrument team could carry out all observations on site. In the case of IGRINS, it would also provide useful staff training for the forthcoming "IGRINS-2" facility instrument that will be contributed by KASI in the future.

Finally, for instruments such as GIRMOS that are built specifically for Gemini, conversion of the instrument from "visiting" to "facility" would be the logical outcome. In this case, Gemini would assume responsibility for the instrument, including operations and maintenance, with a possible compensation package for the instrument team. The instrument would then remain at Gemini indefinitely, providing the user community with stable long-term benefits.

## 8 Capabilities Timeline and Milestones

The preceding sections of this SSP have described specific strategies for how Gemini Observatory plans to reach the "preferred future" envisioned by the Gemini Board in their *Beyond 2021* report. The plan itself is essentially a collection of projects directed towards the overarching goal of increasing Gemini's scientific impact in the coming decade. The graphic timelines in Figure 2 provide a visual summary of the planned progressions of the key projects in this SSP.

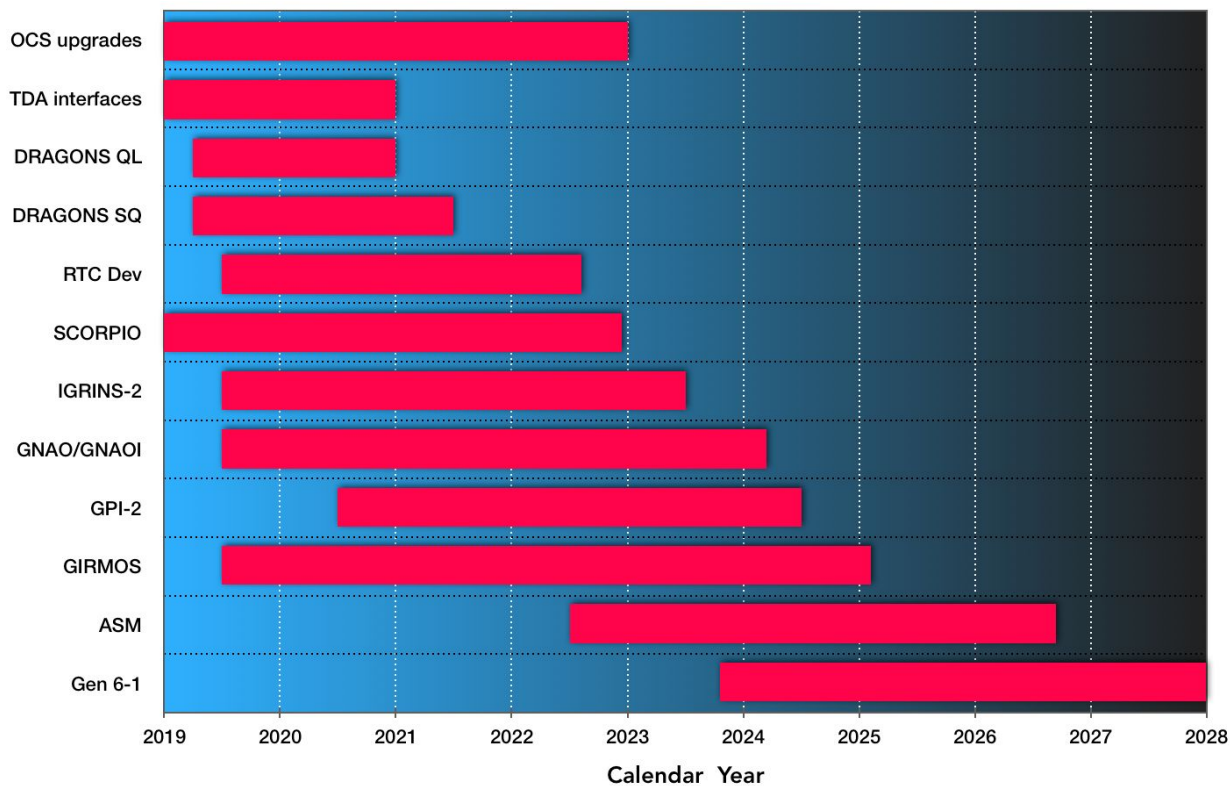

**Figure 2.** Graphical summary of the timelines for the development of the main projects discussed in this SSP; the start dates are not rigorously defined in many cases. The abbreviations QL and SQ refer to "quick look" and "science quality" DRAGONS long-slit spectroscopic data reduction pipelines. Note that IGRINS-2 and GIRMOS are being developed as visiting instruments with the intention of acquiring facility status. Each of the bars in the chart corresponds to an ongoing project at the time of this writing, except for the last three. GPI-2 is not yet funded, but the GPI instrument team is developing plans and pursuing funding opportunities. The STAC has advocated strongly for an adaptive secondary mirror at Gemini North, but because of the numerous other development efforts competing for Observatory resources, the ASM has not yet been formalized as a project in the Gemini portfolio. The final item, denoted "Gen 6-1," represents the first of the next generation of instruments, for which planning should begin in the first quarter of the 2020s with a completion late in the decade.

The timeline illustrated in Figure 2 is aggressive in order to maintain competitiveness throughout the coming decade; the Observatory is committed to the success of these development efforts. In addition, the GEMMA-funded projects have a strict schedule for completion because of the limited six-year term of the award. Adherence to this plan will result in a broad range of new capabilities for the Gemini user community by the middle of the next decade. Of course, to make room for the new instruments that will deliver these capabilities, some of the current facility instruments will need to be retired, as indicated by the list of current and future instruments shown in Table 1 of Section 2 above. In addition, as discussed in Section 7, the Observatory will likely need to be more selective in which visiting instruments it agrees to host. These choices should be driven by community demand for the capabilities, operational impact, and the scientific productivity of each instrument.

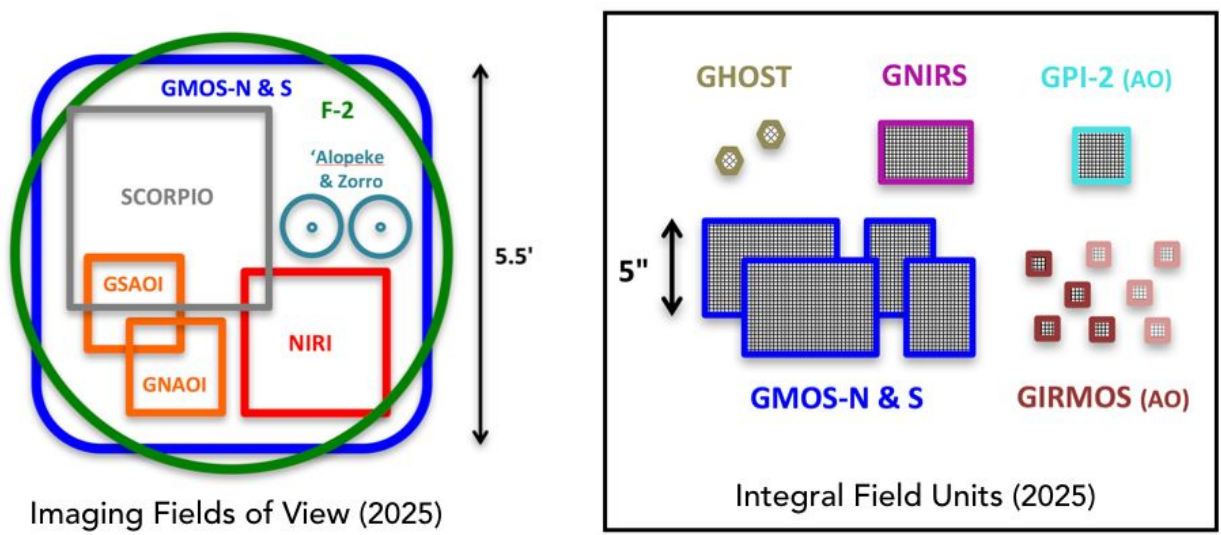

**Figure 3.** Left: Fields of view for Gemini facility and resident imagers in the middle of the coming decade. For simplicity, GNAOI, a generic imager for use with GNAO, is shown as a copy of GSAOI; the capability could also be supplied by an imaging mode of GIRMOS or even an upgraded GSAOI itself. As indicated in Table 1, NIRI will likely need to be retired after GNAO is commissioned, but it could be made available on a "resident" basis if demand is sufficient. Right: Representation of the approximate field sizes of IFUs at Gemini in the same era. NIFS and the AO-assisted GNIRS IFU will likely not be available after GIRMOS is commissioned with GNAO (the preferred option in this SSP). The baseline for GIRMOS is four IFUs, upgradable to eight (indicated by the fainter outlines).

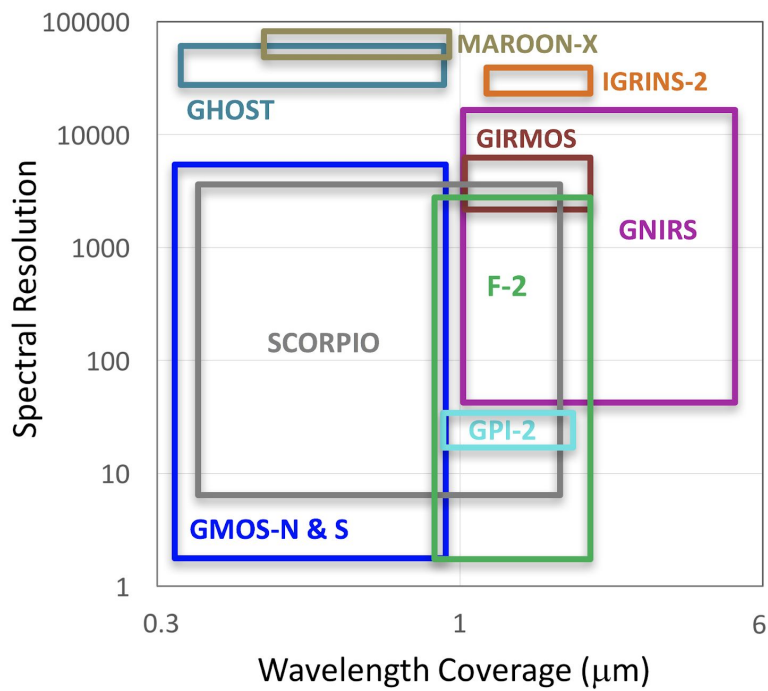

**Figure 4.** Schematic plot of wavelength coverage versus spectral resolution for facility and resident instruments anticipated to be available at Gemini in the mid-2020s. Only instruments with some spectroscopic capability are shown, and the resolution is extended towards ~ 1 if they are frequently used for imaging as well. In the case of SCORPIO, which provides eight bands simultaneously, the effective spectral resolution in imaging mode is higher. Here, GPI-2 is assumed to have the same spectroscopic characteristics as GPI currently does.

The imaging and IFU field sizes of the instruments that may be available at Gemini circa 2025, contingent upon resources and demand, are illustrated in Figure 3. As noted in the caption, GNAOI is shown as a copy of GSAOI, but its characteristics and modes have not yet been fully determined. NIRI will likely be decommissioned as a facility instrument by this time, but could be available for visiting blocks if GNAOI lacks a wider field natural-seeing mode. GIRMOS will also have some imaging capability. Figure 4 gives a schematic view of the spectral resolution and wavelength coverage of the instruments with spectroscopic modes; the figure shows that GNIRS provides a unique long-wavelength capability, especially useful for high-redshift objects, that should be maintained.

Below we provide a total of 23 specific milestones for the 2020s in the three broad areas that fall within the scope of this document: instrumentation, operations, and user support. As of this writing, the Gemini community is anticipating several major milestones in the nearer term, foremost among them being the commissioning of GHOST, which will bring high-dispersion optical spectroscopy to Gemini South and replace GPI in the instrument suite. In addition, the commissioning of NGS-2 will substantially improve the sky coverage of GeMS in the near future, and the visiting instrument MAROON-X will provide high-dispersion spectroscopy and precision velocimetry at Gemini North. These milestones should be reached by or near 2020, and we omit them below. The lists focus on strategic developments that will enable new science opportunities; the associated dates represent the current goals.

**Instrumentation milestones**

1. New GNIRS natural seeing and AO-assisted IFUs commissioned at Gemini North (2021).
2. New RTC commissioned for GeMS at Gemini South, improving efficiency of observations (2022).
3. SCORPIO commissioned at Gemini South (2023).
4. GeMS operable from the Base Facility, enabling more frequent scheduling (2023).
5. IGRINS-2 commissioned, likely at Gemini North (2023).
6. Gemini North Multi-Conjugate Adaptive Optics System commissioned, bringing wide-field diffraction-limited AO capabilities to Maunakea (2024).
7. GPI-2 commissioned at Gemini North (2024).
8. GIRMOS commissioned, likely at Gemini North (2025).
9. Adaptive secondary mirror installed, enabling even wider field corrected images at Gemini North, well-matched to the narrow slit of IGRINS-2 and multi-object spectroscopy with a recommissioned FLAMINGOS-2 or similar instrument (c. 2027).

**Operations milestones**

1. Initial deployment of first components of the upgraded Observatory Control System, including the new database infrastructure and interfaces (APIs) (2020).
2. Automated Queue Planning implemented in advance of AEON scheduler (2021).
3. Demonstration of Gemini North & South being able to operate as nodes on the AEON transient follow-up network via Target Observation Managers (2021).
4. Full incorporation (of participant-determined fraction) of Gemini into AEON (2021).
5. More flexible North-South participant accounting adopted so individual communities can make better use of the capabilities they require; also simplifies dynamic scheduling (2022).

6. Gemini time accessible to PIs through AEON follow-up network proposals (2022).

7. Web-based system to streamline proposing (Phase 1) and designing (Phase 2) observation programs fully integrated into the upgraded OCS software (2023).

8. Start of LSST science operations, with Gemini ready for automated follow-up of brokered transients (2023).

**User Support Milestones related to Data Reduction**

1. Release of automated "quick look" DRAGONS data reduction pipeline to support rapid processing of long-slit spectroscopic follow-up prior to LSST operations (2021).

2. Release of automated "science quality" data reduction pipeline for long-slit spectroscopy (2022).

3. SCORPIO delivered with DRAGONS-based data reduction pipeline (2022).

4. DRAGONS release to support GMOS MOS mode (2023).

5. All Gemini IRAF packages for active facility instruments fully replaced (2024).

6. GIRMOS commissioned with a robust automatic data reduction pipeline (2025).

We realize that these milestones, and the concomitant schedule illustrated in Figure 2, are extremely ambitious, but they are not unrealistic. In striving after and achieving these milestones, Gemini will be furthering the scientific missions of its twin telescopes in Hawaii and Chile by enabling new, cutting-edge science for our diverse user community and improving synergies with major facilities such as LSST, LIGO/Virgo, JWST, TESS, and many others. Sometime in the first half of the coming decade, Gemini should issue a new Request for Proposals for the next generation facility instrument, referred to above as Gen 6-1. Instrument teams could be invited to present their concepts at the 2023 Gemini Science Meeting, and a decision should be made based on the merits of the proposals and the need to continue maximizing the scientific impact of Gemini into the extremely large telescope era of the 2030s.

# 9 Synthesis

Gemini Observatory has a clear path forward for the coming decade. The SSP presented here has been developed based on results from a large survey of the Gemini User Community and recommendations from a variety of Community and Governance committee reports. By following the strategies set forth in this document, Gemini aims to achieve the following objectives:

- Ensuring the current facility instruments are well maintained and instruments now under development are delivered in a timely way;
- Maintaining a diverse variety of proposal options for users to access Gemini time;
- Accommodating the expected larger number of transient programs through a competitive proposal-based time allocation process;
- Rigorously protecting programs that target non-transient sources to prevent any decrease in their completion rates;
- Enhancing operations and participating in the AEON follow-up network, including a dynamic scheduler to optimize the observing plan as conditions and priorities change, while avoiding unnecessary interrupts that would necessitate re-observations;
- Providing robust, automatic data reduction pipelines based on modern software toolkits;
- Completing the development and commissioning of the recently funded, state-of-the-art multi-conjugate AO system to revitalize the mission of Gemini North;
- Deployment of an adaptive secondary mirror at Gemini North to enable GLAO correction for improved performance at all wavelengths, with a timeline contingent on resources;
- Bringing new, in-demand capabilities to the Gemini users through innovative facility-class visitor instruments.

The ultimate goal of this SSP is a more capable, efficient, science-driven Gemini Observatory that serves a broad community of users with diverse scientific interests while excelling in areas for which Gemini's telescopes and operations are especially suited, namely high spatial resolution and time domain science. The realization of this goal will secure Gemini a position at the forefront of astronomy in the coming decade and beyond. There are many ways in which this Plan may fail, and essentially only one by which it can succeed. In order for Gemini to be a success throughout the 2020s, we must move forward together, ensuring that the partnership remains strong, by respecting and valuing the science and contributions of all members: Argentina, Brazil, Canada, Chile, Hawaii, Korea, and the United States.

Strategic Scientific Plan (SSP) for Gemini Observatory    30## Document Authorship and Acknowledgements

The primary author of this *Strategic Scientific Plan for Gemini Observatory* is J. P. Blakeslee, with content and other key contributions provided by:

A. Adamson, C. Davis, R. Díaz, B. Miller, A. Peck, R. Rutten, G. Sivo, J. Thomas-Osip, T. Boroson, R. Carrasco, E. Dennihy, M. Díaz, L. Ferrarese, R. Green, P. Hirst, N. Hwang, I. Jørgensen, H. Kim, S. Kleinman, K. Labrie, T. Lee, J. Lotz, S. Leggett, L. Medina, A. Nitta, J. Pollard, H. Roe, F. Rantakyro, G. Rudnick, R. Salinas, M. Sawicki, and M. Van Der Hoeven.

Gemini Observatory is operated by the Association of Universities for Research in Astronomy, Inc., under a cooperative agreement with the NSF on behalf of the Gemini partnership: the National Science Foundation (United States), National Research Council (Canada), CONICYT (Chile), Ministerio de Ciencia, Tecnología e Innovación Productiva (Argentina), Ministério da Ciência, Tecnologia e Inovação (Brazil), and Korea Astronomy and Space Science Institute (Republic of Korea).## References

Bally, J., Ginsburg, A., Silvia, D., et al. 2015, A&A, 579, A130

Blakeslee, J. P., Rodney, S. A., Lotz, J. M., et al. 2019, BAAS, 51, 529

Chilcote, J. K., Bailey, V. P., De Rosa, R., et al. 2018, Proc. SPIE, 10702, 1070244

Eisner, J. A., Bally, J. M., Ginsburg, A., et al. 2016, ApJ, 826, 16

Fritz, T. K., Linden, S. T., Zivick, P., et al. 2017, ApJ, 840, 30

Hibon, P., Garrel, V., Neichel, B., et al. 2016, MNRAS, 461, 507

Hinz, P. M., Downey, E., Montoya, O. M., et al. 2018, Proc. SPIE, 10703, 1070369

Kelly, P. L., Rodney, S. A., Treu, T., et al. 2015, Science, 347, 1123

Kool, E. C., Ryder, S., Kankare, E., et al. 2018, MNRAS, 473, 5641

Lacy, M., Nyland, K., Mao, M., et al. 2018, ApJ, 864, 8

Macintosh, B., Graham, J. R., Barman, T., et al. 2015, Science, 350, 64

Macintosh, B., Chilcote, J. K., Bailey, V. P., et al. 2018, Proc. SPIE, 10703, 107030

Rigaut, F., Neichel, B., Boccas, M., et al. 2014, MNRAS, 437, 2361

Saracino, S., Dalessandro, E., Ferraro, F. R., et al. 2016, ApJ, 832, 48

Schneider, T., & Stupik, P. 2018, Proc. SPIE, 10700, 1070048

Sivanandam, S., Chapman, S., Simard, L., et al. 2018, Proc. SPIE, 10702, 107021J

Tallis, M., Bailey, V. P., Macintosh, B., et al. 2018, Proc. SPIE, 10703, 1070356

Tegler, S. C., Stufflebeam, T. D., Grundy, W. M., et al. 2019, AJ, 158, 17